# Integration of FogBus2 Framework with Container Orchestration Tools in Cloud and Edge Computing Environments

by

Zhiyu Wang

Supervised under Prof. Rajkumar Buyya

A project report submitted for the

25-pts Research Project COMP90055

and degree of Master of Information Technology (Artificial Intelligence)

in the

School of Computing and Information Systems

The University of Melbourne, Australia

October 2021

# Integration of FogBus2 Framework with Container Orchestration Tools in Cloud and Edge Computing Environments

Zhiyu Wang

*Supervisor: Prof. Rajkumar Buyya*

# Abstract

Currently, due to the advantages of light weight, simple deployment, multi-environment support, short startup time, scalability, and easy migration, container technology has been widely used in both cloud and edge/fog computing, and addresses the problem of device heterogeneity in different computing environments. On this basis, as one of the most popular container orchestration and management systems, Kubernetes almost dominates the cloud environment. However, since it is primarily designed for centralized resource management scenarios where computing resources are sufficient, the system is unstable in edge environments due to hardware limitations. Therefore, in order to realize container orchestration in the cloud and edge/fog hybrid computing environment, we propose a feasible approach to build a hybrid clustering based on K3s, which solves the problem that virtual instances in different environments cannot be connected due to IP addresses. We also propose three design patterns for deploying the FogBus2 framework into hybrid environments, including 1) Host Network Mode, 2) Proxy Server, and 3) Environment Variable.

# Declaration

I certify that

1. This project report does not include any other individual or group scientific research results that have been published or written except for the content cited in the text.

2. The project report is less than 7000 words in length, exclusive of text in figures, tables, and bibliographies.

___________________________________

Zhiyu Wang, October 2021

# Acknowledgments

First of all, I would like to thank my supervisor, Professor Rajkumar Buyya. From the initial selection of my research direction to the finalization of my project report, Professor Buyya has provided friendly guidance and support. His professionalism, rigorous academic attitude and dedication to excellence have profoundly influenced and inspired me. I would like to express my sincere gratitude and respect to Professor Buyya.

I also want to express my gratitude to Mr. Mohammad Goudarzi, who has provided me with plenty of help and suggestions during the implementation of my project. His selfless support and encouragement, and rich experience guided me to complete this project successfully.

Finally, I want to thank my family for their unconditional love and support for me materially and morally. Their support and care not only give me spiritual inspiration, but also become an inexhaustible motivation for my continuous progress.

# Contents

# List of Figures

# List of Tables

# Chapter 1  Introduction

With the rapid development of network technology, IoT devices have penetrated into all aspects of our lives. Traditional physical devices are connected in the Internet of Things environment to perform anthropomorphic information perception and collaborative interaction. They realize self-learning, processing, decision-making, and control, thereby completing intelligent production and service, and promoting the innovation of people's life and work patterns [1].

On this premise, cloud computing, with its powerful computing and storage capabilities, becomes a shared platform for IoT big data analysis and processing. In most cases, IoT devices upload complex applications to the cloud for storage and processing, and the output results are then sent from the cloud to end devices [2]. As a result, users no longer need to worry about insufficient storage space or processing speed for IoT end devices. However, with the explosive growth in the number of IoT end devices nowadays, the amount of raw data sensed and acquired by the IoT has been increasing significantly, and there are complicated relationships between the massive amounts of data. Consequently, filtering, processing, and analyzing the massive amount of data has become an inevitable challenge for the cloud computing framework [2].

Moreover, while the Internet of Things is gradually impacting society as a whole, edge computing is becoming a popular solution to empower it. As a computing architecture, edge computing concentrates data and processing as close to the end-user as possible, as opposed to traditional cloud computing architectures that concentrate data and processing in cloud data centers [3]. The key idea behind edge computing is that network latency and reliability are reduced when workloads are hosted closer to the user, resulting in a better end-user experience [4]. However, while edge computing can cope with some everyday medium to lightweight tasks, when the user's needs involve complex computing and resource usage, edge devices are often unable to meet the demanding requirements needed for the task because they have limited computing performance [3, 5].

With such challenges, hybrid computing environments are becoming popular solutions. Edge and cloud computing work in tandem, and optimally complement each other. Cloud computing acts as the role of an orchestrator, which is responsible for big data analysis of long-period data and is able to operate in areas such as cyclical maintenance and business decisions. Edge computing, on the other hand, looks at the analysis of real-time, short-period data to better support the timely processing and execution of local tasks.

## 1.1 Motivation and Challenges

Edge computing has undergone an extremely rapid technological evolution in the past few years, and the combination with the cloud model will enable edge computing to better absorb the results of cloud, big data, and AI, and allow the latter to further extend its application scope [5]. However, in practice, contradicting the strong market demand, edge computing does not yet have a mature technology system, with problems including serious heterogeneity of edge devices, no unified architecture, the large number and wide distribution of edge devices, and lack of technical standards and specifications. Meanwhile, container technology has been developing significantly in recent years. Compared with physical and virtual machines, container technology is very lightweight, simple to deploy, supports multiple environments, has a shorter start-up time, and is easy to expand and migrate. These features are a good solution to the problem of severe heterogeneity of edge devices, and are gradually being used by industry and academia to run commercial, scientific, and big data applications, build IoT and edge/fog computing systems, and manage internal infrastructure and services [6].

However, the increase in the number of containers may make project coordination increasingly difficult. The use of containers, while allowing for a smooth workflow for programmers, cannot be automatically applied to production environments. In addition, managing large clusters of containers and nodes can add to the burden of developers. In this environment, the introduction of container orchestration technologies to automate the management of application deployment, scalability, and network in cluster environments has become a popular practice.

While container orchestration tools such as Kubernetes have become the ideal solution for managing and scaling deployments, nodes, and clusters in the industry today [7], there are still a number of challenges with their application in hybrid cloud and edge/fog environments. Firstly, orchestration techniques need to take into account the heterogeneity in computing resources of different environments in order to achieve wide adaptability. Secondly, the complexity of installing and configuring hybrid cloud and edge/fog environments needs to be addressed when implementing orchestration techniques. Thirdly, a strategy needs to be investigated to solve potential conflicts between orchestration techniques and the network model of the hybrid environment.

## 1.2    Research Problem

This project report explores the feasibility of deploying container orchestration tools in cloud and edge/hybrid environments to enable resource limit control, health checks, and fault tolerance for containers. Specifically, to address the previously mentioned challenges, the following questions should be answered in the study.

- **Can the deployment of container orchestration tools be implemented in a hybrid cloud and edge/fog environment?**

    In a hybrid cloud and edge/fog environment, the collaborative expression of different containers can be abstracted into applications for the users of the cluster environment. In this case, deploy container orchestration is similar to local implementation. However, mainstream container orchestration tools usually have high resource requirements for computer hardware devices, although cloud

instances are often powerful in terms of computing resources, edge devices are constrained by cost and market factors that often prevent them from being well loaded with these container orchestration tools. As a result, container orchestration tools suitable for edge environments should be used to ensure that containers running on both the cloud and edge can be managed efficiently.

- **How can container orchestration tools be deployed in the complex challenges of hybrid edge/fog networks?**

    Hybrid cloud and edge/fog environments require a dedicated network environment to enable different instances and applications to communicate with each other, given the complexity of the network. Container orchestration tools, when deployed in clusters, typically create their own container-based network environment to enable resource scheduling and container fault tolerance. In order to implement container orchestration in hybrid cloud and edge/fog environments, a unified network environment or mutually compatible policies need to be investigated to ensure that the two can work in harmony.

## 1.3 Methodology

The objective of this project report is to deploy container orchestration technology into the FogBus2 framework [8] to automate container management. Compared to Fogbus2, the new system allows for the implementation of resource limit control, health checks, and fault tolerance to cope with the changing number and functionality of IoT devices connected. To enable container orchestration for hybrid cloud and edge/fog environments, the following methodologies were used.

- **Hybrid clusters:** To implement container orchestration techniques in the FogBus2 framework, we needed to build a cluster for the hybrid edge/fog and cloud computing environment. In practice, we used K3s, a lightweight Kubernetes distribution that is highly optimized for edge computing, IoT, and other scenarios [9], to connect instances in the cloud to edge devices. For the actual implementation, we used three Nectar cloud instances to simulate the cloud environment and created two Linux virtual machines to simulate the edge devices. K3s was able to run smoothly in both environments.

- **Network connection:** Under normal circumstances, IoT devices and edge devices provide services in a local area network without public IP addresses. In order to simulate this situation, we did not set a public IP address on the Linux virtual machines, and this posed a challenge for cluster creating, because K3s usually uses public IP addresses to connect to different hosts. To cope with this problem, we deployed virtual private networks on cloud instances and edge virtual machines to achieve interconnection between all hosts.

- **Communication strategy:** The Fogbus2 framework communication model is designed in such a way that components built on different hosts need to use host IP addresses to transfer information, and the network planning service within the K3s cluster will assign a cluster-wide unique virtual IP address to each container created by hosts on different nodes, which causes compatibility issues when deploying the Fogbus2 framework in the cluster. Furthermore, while virtual private networks can be used to create edge/fog and cloud hybrid environments cluster, it will affect the network configuration of the K3s, which reflects in practice that applications located on different nodes cannot communicate with each other. Therefore, to address these issues and to enable the Fogbus2 framework to run smoothly in K3s, we use the Host Network mode to deploy applications in clusters.

## 1.4   Project Report Contributions

Based on the discussed research problem of deploying container orchestration techniques in cloud and edge/fog computing environments, the main contributions of this project report can be summarized as follows:

- Present feasible designs for implementing container orchestration techniques in cloud and edge/fog computing environments.

- Solve the problem of connecting virtual instances due to IP addresses in different environments by building a hybrid environment with cloud nodes and edge devices using VPNs.

- Propose three design patterns for deploying the FogBus2 framework into the hybrid environment.

## 1.5   Project Report Organization

The rest part of this project report is organized as follows. Chapter 2 provides a background on the relevant technologies and reviews the deployment of container orchestration technologies in cloud and fog environments. Chapter 3 describes the creation of the K3s cluster and the details of the implementation in deploying the FogBus2 framework into the K3s cluster. Chapter 4 concludes the project report and presents future works. A visual organization of the project report is shown in Figure 1.1.

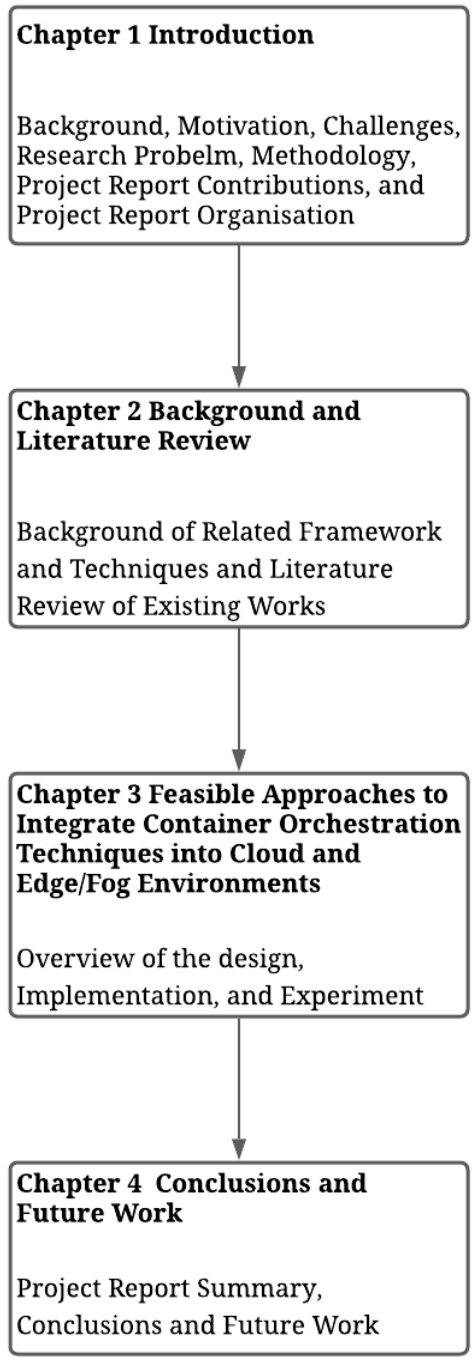

**Figure 1.1:** Visualized Project Report Organization

# Chapter 2 Background and Literature Review

This chapter discusses the framework and container orchestration techniques involved in the project, including the FogBus2 framework, Kubernetes, K3s, and Minikube. In addition, this chapter also reviews the research on container orchestration in the cloud and edge/fog environments.

## 2.1 Background of Related Framework and Techniques

### 2.1.1 FogBus2 Framework

FogBus2 is a lightweight distributed container-based framework developed by the CLOUDS Laboratory, University of Melbourne. FogBus2 integrates edge/fog and cloud infrastructures to sup FogBus2.port the execution of heterogeneous IoT applications [10]. Figure 2.1 shows an overview of the computing environments supported by FogBus2.

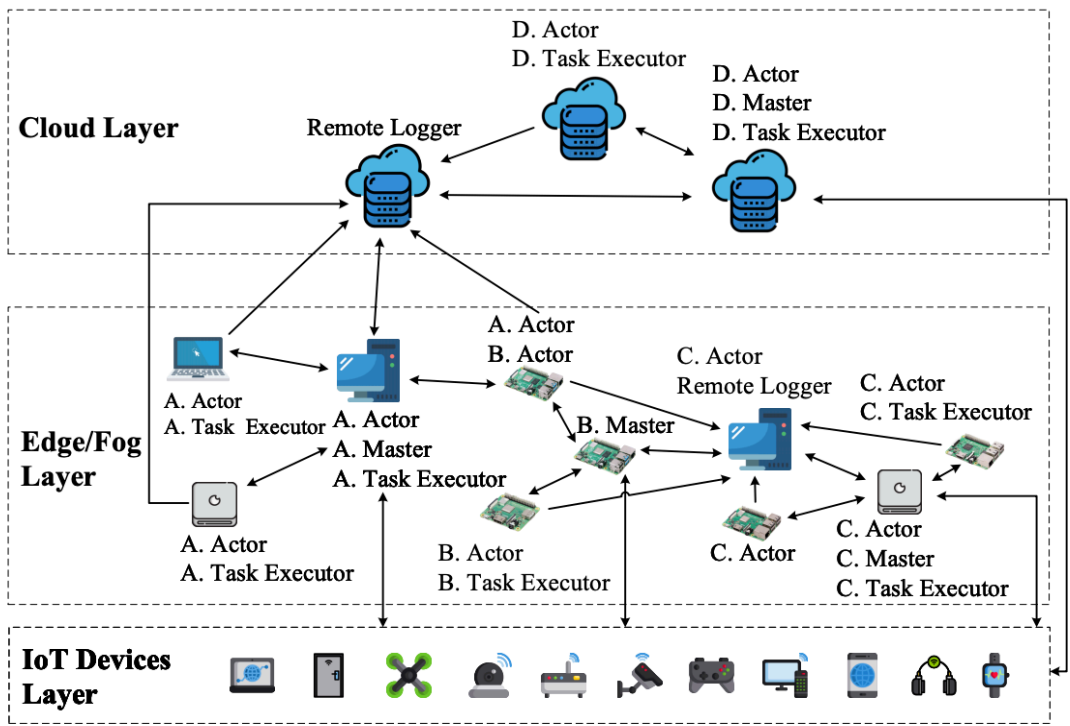

**Figure 2.1:** Overview of the computing environments supported by FogBus2 [8]

## 2.1.1.1 Hardware Architecture

From a hardware perspective, the FogBus2 framework consists of an IoT device layer, an edge/fog layer, and a cloud layer [8, 10]. The IoT device layer contains various types of resource-constrained IoT devices that sense data from the environment and perform physical operations and can transmit the generated results to higher-level proxy servers; the edge/fog layer includes devices such as Raspberry Pi (RPi), personal computers, routers, and gateways that provide low-latency and high-bandwidth services to IoT devices; the cloud layer extends the compute and storage resources of IoT devices and can be used to address heavy tasks when the edge/fog layer resources are overloaded [8].

## 2.1.1.2 Software Components

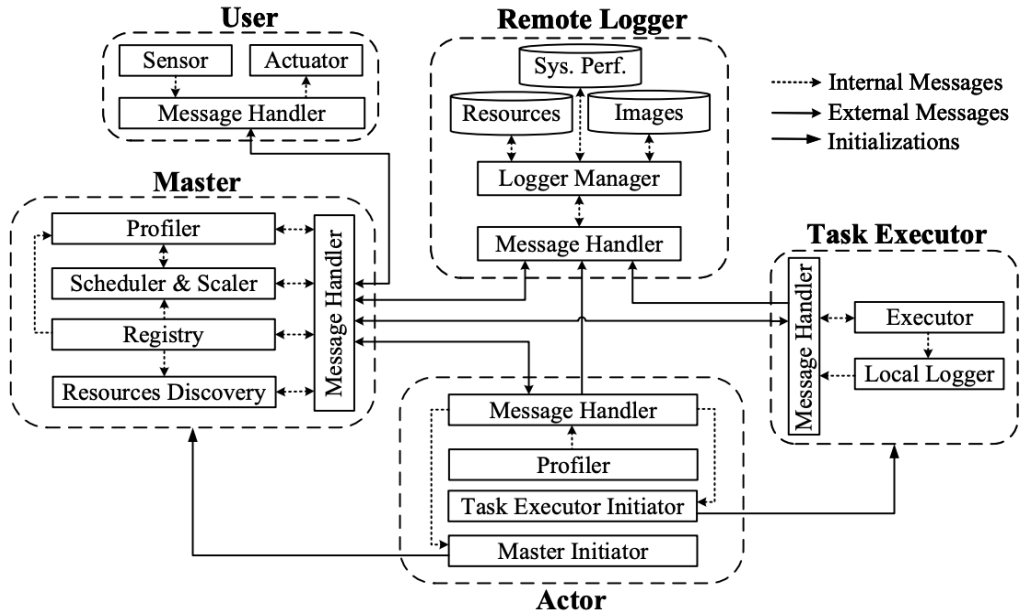

**Figure 2.2:** Software components and interaction model of FogBus2 [10]

From a software perspective, FogBus2 contains five main components developed in Python that run in a docker container [8], so that they can be deployed on different hosts depending on the application scenario. Figure 2.2 shows the software components and interaction model of FogBus2.

- **User:** The User component runs on the user's IoT device to receive raw data from sensors and send placement requests to the Master component [8, 10].

- **Master:** The Master component can run on any host in the edge/fog or cloud layer depending on the application scenario, handling placement requests from IoT devices and managing the execution of IoT applications. In addition, it can dynamically analyze the environment and find available compute and storage resources [8, 10].

- **Actor:** The Actor component can run on any host in the edge/fog or cloud layer to receive task commands from the Master component and initiate the appropriate

Task Executor component. Besides, it can be transformed into a Master component under certain conditions to achieve architectural scalability [8, 10].

- **Task Executor:** The Task Executor component is used to execute the tasks of the application and can be efficiently reused for other requests of the same type, thus significantly reducing the deployment time of the task [8, 10].

- **Remote Logger:** The Remote Logger component can run on any host in the Edge/Fog or Cloud layers, collecting periodic or event-driven logs from other components and storing them in persistent storage using a file system or database [8, 10].

## 2.1.2 Container Orchestration Techniques

There are many container orchestration tools such as Kubernetes, K3s, and Minikube with different properties. In what follows, we discuss three of these container orchestration tools in detail.

### 2.1.2.1 Kubernetes

Kubernetes is an open-source container cluster management system based on container technology [11]. It provides a series of complete functions such as deployment and operation, resource scheduling, service discovery, and dynamic scaling for containerized applications, which improves the convenience of large-scale container cluster management [11]. Figure 2.3 shows an overview of the Kubernetes architecture.

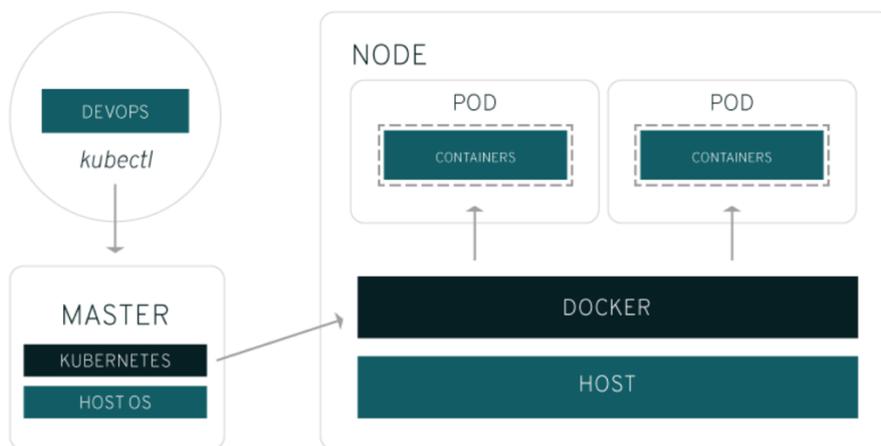

**Figure 2.3:** An overview of the Kubernetes architecture [12]

- **Master:** The management node of Kubernetes, responsible for managing the cluster, and providing access to the cluster's resource data [12].

- **Node:** The unit of Kubernetes cluster operation, used to carry the operation of the assigned Pod, and is the host machine of the Pod operation [12].

- **Pod:** It runs on Nodes and contains a combination of one or more related containers. The containers contained in the Pod run on the same host, use the same network namespace, IP address, and port, and can communicate through localhost. A pod is the smallest unit created, scheduled, and managed by Kubernetes. It provides a higher level of abstraction than containers, making deployment and management more flexible [11, 12].

### 2.1.2.2 K3s: Lightweight Kubernetes

K3s is a lightweight Kubernetes designed for environments with limited resources, suitable for IoT, edge computing, and ARM devices [9]. Compared to Kubernetes, K3s is half the size in terms of memory footprint, but API consistency and functionality are not compromised [9]. Figure 2.4 shows the architecture of a K3s cluster containing one

Server and Multiple Agents.

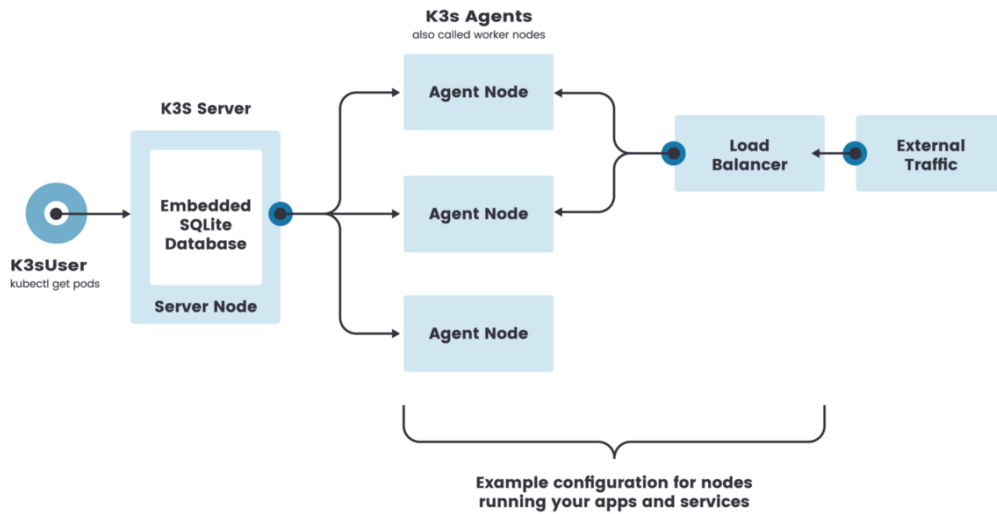

**Figure 2.4:** The architecture of a single Server K3s cluster [13]

K3s clusters allow Pods to be scheduled and managed on any node. Similar to Kubernetes, K3s clusters also contain two types of nodes, with the Server running the control plane components and kubelet, and the Agent running only the kubelet [13]. Typically a K3s cluster carries a Server and multiple Agents. When the URL of a Server is passed to a K3s node, that node becomes an Agent; otherwise it is a Server in a separate K3s cluster [9, 13].

### 2.1.2.3 Minikube

Minikube is a standalone Kubernetes cluster maintained by the Kubernetes community [14]. It runs on a variety of operating systems including macOS, Linux, and Windows, and supports most of the features of Kubernetes, from basic container orchestration management to advanced features such as permission control, load balancing, Ingress, etc., suitable for use as an introduction to Kubernetes, or as a development and testing environment [14].

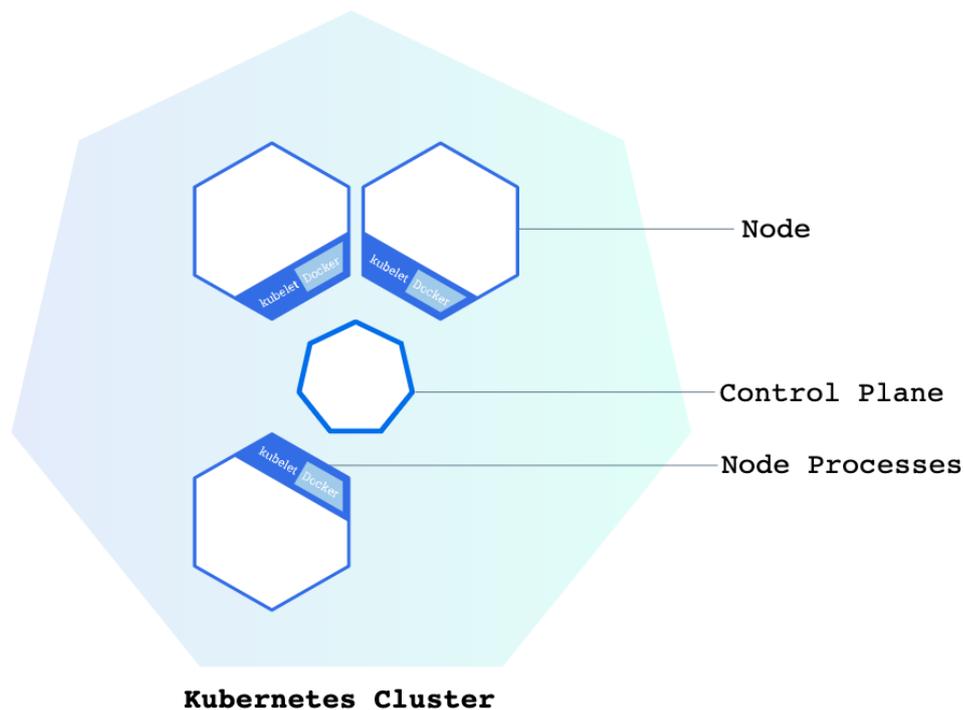

**Figure 2.5:** The diagram of a Minikube Kubernetes cluster [15]

However, Minikube only supports the creation of single-node Kubernetes clusters, and multi-node applications are still being planned [15].

## 2.2 Literature Review of Existing Works

As a lightweight and distributed container-based framework, FogBus2 developed by Deng al. [8] integrates edge and cloud environments to implement multiple scheduling policies for scheduling heterogeneous IoT applications. In addition, it proposes an optimized genetic algorithm for fast convergence of resource discovery to implement a scalable mechanism that addresses the problem that the number of IoT devices increases or resources become overloaded. Besides, the dynamic resource discovery mechanism of FogBus2 facilitates the rapid addition of new entities to the system. The work of Rodriguez al . [16] investigates multiple container orchestration platforms and proposes a taxonomy of different mechanisms that can be used to cope with fault tolerance, availability, scalability, efficient resource utilization and maximization of request throughput, etc. Zhong al . [17] proposed a Kubernetes-based container orchestration strategy for cost-effective container orchestration in cloud environments. It supports heterogeneous job configuration, cluster resizing, and rescheduling mechanisms for resource utilization, optimization, and elastic instance pricing.

Furthermore, FLEDGE developed by Goethals al. [18] implements container orchestration in an edge environment that is compatible with Kubernetes, but with 50-60% less node resource usage than Kubernetes. Pires al. [19] proposes a framework named Caravela that employs a fully decentralized architecture, resource discovery, and scheduling algorithms. It leverages users' voluntary edge resources to build an independent environment where applications can be deployed using standard Docker containers to cope with large numbers of voluntary devices, unstable environments, wide area networks of connected devices, and the absence of central management. Alam el. [20] proposed a layered modular architecture running on the cloud, fog, and edge devices, providing containerized services and microservices. Based on lightweight virtualization, it creates a highly dynamic system by combining modularity with the orchestration provided by Docker, enabling distributed deployment and simplified management. Ermolenko el. [21]'s work studied a framework for deploying IoT applications based on Kubernetes orchestrator and microservice manager in the edge cloud environment. It achieves lightweight scaling of task-based applications and allows the addition of external data warehouses.

# Chapter 3  Feasible Approaches to Integrate Container Orchestration Techniques into Cloud and Edge/Fog Environments

Based on the research on the FogBus2 framework and container orchestration tools, as well as the review of existing work, we propose a feasible approach for deploying Container Orchestration Techniques in Cloud and Edge/Fog Hybrid Environments. Chapter 3.1 presents a high-level overview of the design. Chapter 3.2 introduces the concrete implementation of the proposed approach. Chapter 3.3 demonstrates the experimental procedures and evaluates the differences between the FogBus2 framework running in the K3s cluster and in the native environment in terms of the response time.

## 3.1   Overview of the Design

To fit the context of the FogBus2 architecture, we need to create a cloud and edge/fog hybrid environment. We chose K3s as the basis for the hybrid environment, because Kubernetes requires high computing resources and is commonly used for large-scale cluster deployments. In contrast, K3s only occupies less than half of the resources of Kubernetes, and is specially optimized for the edge computing environment, suitable in resource-constrained scenarios. Minikube only supports the creation of one master node cluster, which sacrifices some Kubernetes functions and is mainly for learning and building test environment, while K3s fully implements the Kubernetes API, which is more suitable for practical engineering applications. In practice, we used three Nectar instances to simulate the cloud environment and created two Linux virtual machines on a physical host to simulate the edge layer. However, our Cloud nodes have public IP addresses, while Edge nodes do not hold public IP addresses. To address this problem, we used Wireguard to set up a lightweight P2P VPN connection among all servers. The hybrid clustering environment was now complete and we started to embed the FogBus2 architecture into it. To take advantage of the container orchestration tool, we allocated only one container to each Pod created by K3s, with only one component of the FogBus2 architecture running inside each container. Also, to balance the load on each node between clusters, we assigned all Pods to different nodes. However, during our experiments, we found that the initialization of the FogBus2 components required binding the host IP address, which would be used to pass information between the different components. This means that in K3s clustering, the FogBus2 component needs to bind the IP address of the Pod, which poses a difficulty for the implementation, as usually the Pod is created at the same time as the application is deployed. To address this problem, we evaluated three approaches and finally decided to use host network mode to deploy the FogBus2 framework in K3s hybrid clustering. Host network mode allows Pods to use the network configuration of virtual instances or physical hosts directly, which solves the communication problem of the FogBus2 components and the conflict between K3s network planning service and VPN. Figure 3.1 shows a high-level overview of our proposed design pattern.

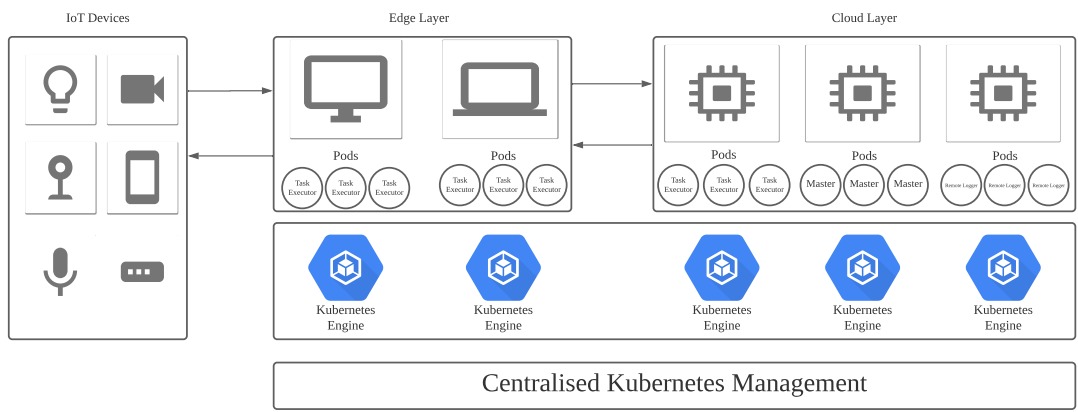

**Figure 3.1:** Overview of the design pattern

## 3.2 Implementation

This section describes the specific process of project implementation, the difficulties encountered, and the coping strategies. The content includes the establishment of VPN, the construction of K3s cloud and edge/fog hybrid environment, and the integration of FogBus2.

### 3.2.1 Configuration of Nodes in Integrated Computing Environments

Our hybrid integrated computing environment consists of five instances, labeled A through E. We use three Nectar instances to set up the cloud environment. In addition, we used two Ubuntu virtual machines on Mac OS to build an edge computing layer with heterogeneous resources. The server list, computing layer, public network IP address, and private network IP address after the VPN connection is established are shown in Table 3.1.

| Server Tag | Server Name | Computing Layer | Public IP Address | Private IP Address | Port | Environment Preparation |
|---|---|---|---|---|---|---|
| A | Nectar1 | Cloud | 45.113.235.156 | 192.0.0.1 | automatically assign | docker |
| B | Nectar2 | Cloud | 45.113.232.199 | 192.0.0.2 | automatically assign | docker |
| C | Nectar3 | Cloud | 45.113.232.232 | 192.0.0.3 | automatically assign | docker |
| D | Virtual Machine1 on a Desktop | Edge | - | 192.0.0.4 | automatically assign | docker |
| E | Virtual Machine2 on a Desktop | Edge | - | 192.0.0.5 | automatically assign | docker |

**Table 3.1:** Configuration of Nodes in Integrated Computing Environment

## 3.2.2 P2P VPN Establishment

As shown in Table 3.1, Cloud nodes have public IP addresses, while in most cases, devices in the Edge environment obtain their IP addresses from DHCP servers and do not have public IP addresses. In this case, in order to build a hybrid cluster, we need to establish a VPN connection to integrate the Cloud and Edge nodes. We used Wireguard to establish a lightweight P2P VPN connection between all the servers. In the implementation, we installed the Wireguard tool for each node and generated the corresponding configuration scripts to ensure that each node had direct access to all other nodes in the cluster. A typical configuration script for Wireguard VPN is shown in Figure 3.2.

```
# *** For worker04 Only ***

[Interface]
PrivateKey = qA+AhTAA+Y5MQFW8tQ/3YbgH3XvNo3VDlba3wlyiLnM=
Address = 192.0.0.5/24
ListenPort = 4999

[Peer]
# master
PublicKey = sZgxVFEs04zxIA9N0dWIo7SHn5vHMIgrRGOk0tkpCnE=
Endpoint = 45.113.235.156:4999
AllowedIPs = 192.0.0.1/32
PersistentKeepalive = 15

[Peer]
# worker01
PublicKey = JiuDTBe26S9rqPyR0tbTsgNQ7M9+eBM51IsbeuN2gUk=
Endpoint = 45.113.232.199:4999
AllowedIPs = 192.0.0.2/32
PersistentKeepalive = 15

[Peer]
# worker02
PublicKey = VER5lsB6VsTanrEwkVS0DDlBieZnC1Refdnw/P0r7VU=
Endpoint = 45.113.232.232:4999
AllowedIPs = 192.0.0.3/32
PersistentKeepalive = 15

[Peer]
# worker03
PublicKey = 9kSrAu7K4NVoOGk37bGXY7F/m+8XGANX0YlZNuMzu34=
Endpoint = 192.168.0.40:4999
AllowedIPs = 192.0.0.4/32
PersistentKeepalive = 15
```

**Figure 3.2:** A sample configuration script for Wireguard VPN creation

## 3.2.3 Construction of K3s Cloud and Edge/Fog Hybrid Environment

To build the Cloud and Edge/Fog Hybrid Environment, we created five Ubuntu 20 instances, two in the cloud with 9GB of RAM and two VCPUs, and three in the edge tier with 1GB of RAM and one VCPU. One of the instances located in the cloud acts as the master node and the other four as the worker nodes. As the aim of our research is to implement container orchestration on the FogBus2 framework, we need to install and enable Docker on both master and worker nodes before we can build K3s. First, we installed and started the K3s server in Docker mode on the master node. K3s allows users to choose the appropriate container tool, but as all components of FogBus2 run natively in Docker containers, we used Docker mode to initialize the K3s master server to allow the cluster to access the Docker images on the host. Then, we extracted a token from the master node, which will be used to join other nodes to the master node. After that, we installed the K3s server on each worker node, specifying the IP of the master node and the token obtained from the master during installation to ensure that all worker nodes could properly connect to the master node. Figure 3.3 shows the successfully running K3s Cloud and Edge/Fog hybrid cluster.

```
NAME        STATUS    ROLES                  AGE    VERSION
worker04    Ready     <none>                 15d    v1.21.5+k3s1
worker03    Ready     <none>                 15d    v1.21.5+k3s1
worker01    Ready     <none>                 15d    v1.21.5+k3s1
worker02    Ready     <none>                 15d    v1.21.5+k3s1
master      Ready     control-plane,master   15d    v1.21.5+k3s1
```

**Figure 3.3:** K3s Cloud and Edge/Fog Hybrid Environment

### 3.2.4 Fogbus2 Framework Integration

In the native design of the FogBus2 architecture, all components are running in containers and the Pod, as the smallest unit created and deployed by K3s, is an application instance in the cluster. Users can wrap one or more containers into a single Pod. Any containers in the same Pod will share the same namespace and local network. Containers can easily communicate with other containers in the same or different Pod, as if they were on the same machine while maintaining a degree of isolation. So first, we are faced with the choice of assigning only one container per Pod (i.e., a component that the FogBus2 architecture is built on) or allowing each Pod to manage multiple containers. The former would balance the load as much as possible between clusters and reduce coupling between applications to facilitate management by the K3s controller, while the latter would reduce the time taken to communicate between components and provide faster feedback to users. Through evaluation and discussion, we decided to adopt the former solution, as the goal of this project was to improve the efficiency of container orchestration and fault tolerance of the clustering controller as opposed to reducing the communication time.

In order to integrate all types of components in the FogBus2 framework into K3s clustering, we first defined the YAML deployment file for necessary components. This file is used to provide the object's statute, which describes the expected state of the object, as well as some basic information about the object. In our project, the YAML deployment file serves to declare the number of replicas of the Pod, the node it is built on, the name of the image, the image pulling policy, the parameters for application initialization, and the location of the mounted volumes. The following scripts illustrate the YAML deployment file for necessary components of the FogBus2 framework.

```yaml
1.  # YAML deployment file for the Master component of the FogBus2 framework
2.  apiVersion: apps/v1
3.  kind: Deployment
4.  metadata:
5.    annotations:
6.      kompose.cmd: /snap/kompose/19/kompose-linux-amd64 convert --volumes hostPath
7.      kompose.version: 1.21.0 (992df58d8)
8.    creationTimestamp: null
9.    labels:
10.     io.kompose.service: fogbus2-master
11.   name: fogbus2-master
12. spec:
13.   replicas: 1
14.   selector:
15.     matchLabels:
16.       io.kompose.service: fogbus2-master
17.   strategy:
```

```yaml
18.      type: Recreate
19.    template:
20.      metadata:
21.        annotations:
22.          kompose.cmd: /snap/kompose/19/kompose-linux-amd64 convert --volumes hostPath
23.          kompose.version: 1.21.0 (992df58d8)
24.        creationTimestamp: null
25.        labels:
26.          io.kompose.service: fogbus2-master
27.      spec:
28.        containers:
29.        - env:
30.          - name: PGID
31.            value: "1000"
32.          - name: PUID
33.            value: "1000"
34.          - name: PYTHONUNBUFFERED
35.            value: "0"
36.          - name: TZ
37.            value: Australia/Melbourne
38.          image: cloudslab/fogbus2-remote_logger
39.          imagePullPolicy: ""
40.          name: fogbus2-master
41.          args: ["--bindIP", "192.0.0.1", "--bindPort", "5001",
42.                 "--remoteLoggerIP", "192.0.0.1", "--remoteLoggerPort", "5000",
43.                 "--schedulerName", "RoundRobin", "--containerName", "TempContainerName"]
44.          resources: {}
45.          volumeMounts:
46.          - mountPath: /var/run/docker.sock
47.            name: fogbus2-master-hostpath0
48.          - mountPath: /workplace/
49.            name: fogbus2-master-hostpath1
50.          - mountPath: /workplace/.mysql.env
51.            name: fogbus2-master-hostpath2
52.        restartPolicy: Always
53.        serviceAccountName: ""
54.        nodeName: master
55.        hostNetwork: true
56.        volumes:
57.        - hostPath:
58.            path: /var/run/docker.sock
59.          name: fogbus2-master-hostpath0
60.        - hostPath:
61.            path: /home/hehe/FogBus2/containers/master/sources
```

```yaml
62.         name: fogbus2-master-hostpath1
63.       - hostPath:
64.           path: /home/hehe/FogBus2/containers/master/sources/.mysql.env
65.         name: fogbus2-master-hostpath2
66. status: {}
```

```yaml
1.  # YAML deployment file for the Remote Logger component of the FogBus2 framework
2.  apiVersion: apps/v1
3.  kind: Deployment
4.  metadata:
5.    annotations:
6.      kompose.cmd: /snap/kompose/19/kompose-linux-amd64 convert --volumes hostPath
7.      kompose.version: 1.21.0 (992df58d8)
8.    creationTimestamp: null
9.    labels:
10.     io.kompose.service: fogbus2-remote-logger
11.   name: fogbus2-remote-logger
12. spec:
13.   replicas: 1
14.   selector:
15.     matchLabels:
16.       io.kompose.service: fogbus2-remote-logger
17.       run: fogbus2-remote-logger
18.   strategy:
19.     type: Recreate
20.   template:
21.     metadata:
22.       annotations:
23.         kompose.cmd: /snap/kompose/19/kompose-linux-amd64 convert --volumes hostPath
24.         kompose.version: 1.21.0 (992df58d8)
25.       creationTimestamp: null
26.       labels:
27.         io.kompose.service: fogbus2-remote-logger
28.         run: fogbus2-remote-logger
29.     spec:
30.       containers:
31.       - env:
32.         - name: PGID
33.           value: "1000"
34.         - name: PUID
35.           value: "1000"
36.         - name: TZ
```

```yaml
37.            value: Australia/Melbourne
38.          image: cloudslab/fogbus2-remote_logger
39.          imagePullPolicy: ""
40.        ports:
41.        - containerPort: 5000
42.          protocol: TCP
43.          name: fogbus2-remote-logger
44.          args: ["--bindIP", "192.0.0.1", "--containerName", "fogbus2-remote-logger"]
45.          resources: {}
46.          volumeMounts:
47.          - mountPath: /var/run/docker.sock
48.            name: fogbus2-remote-logger-hostpath0
49.          - mountPath: /workplace/.mysql.env
50.            name: fogbus2-remote-logger-hostpath1
51.      restartPolicy: Always
52.      serviceAccountName: ""
53.      nodeName: master
54.      hostNetwork: true
55.      volumes:
56.      - hostPath:
57.          path: /var/run/docker.sock
58.        name: fogbus2-remote-logger-hostpath0
59.      - hostPath:
60.          path: /home/hehe/FogBus2/containers/remoteLogger/sources/.mysql.env
61.        name: fogbus2-remote-logger-hostpath1
```

```yaml
1.  # YAML deployment file for the Actor component of the FogBus2 framework
2.  apiVersion: apps/v1
3.  kind: Deployment
4.  metadata:
5.    annotations:
6.      kompose.cmd: /snap/kompose/19/kompose-linux-amd64 convert --volumes hostPath
7.      kompose.version: 1.21.0 (992df58d8)
8.    creationTimestamp: null
9.    labels:
10.     io.kompose.service: fogbus2-actor
11.   name: fogbus2-actor
12. spec:
13.   replicas: 1
14.   selector:
15.     matchLabels:
16.       io.kompose.service: fogbus2-actor
```

```yaml
17.    strategy:
18.      type: Recreate
19.    template:
20.      metadata:
21.        annotations:
22.          kompose.cmd: /snap/kompose/19/kompose-linux-amd64 convert --volumes hostPath
23.          kompose.version: 1.21.0 (992df58d8)
24.        creationTimestamp: null
25.        labels:
26.          io.kompose.service: fogbus2-actor
27.      spec:
28.        containers:
29.        - env:
30.          - name: PGID
31.            value: "1000"
32.          - name: PUID
33.            value: "1000"
34.          - name: PYTHONUNBUFFERED
35.            value: "0"
36.          - name: TZ
37.            value: Australia/Melbourne
38.          - name: MY_POD_IP
39.            valueFrom:
40.              fieldRef:
41.                fieldPath: status.podIP
42.          image: cloudslab/fogbus2-actor
43.          imagePullPolicy: ""
44.          name: fogbus2-actor
45.          args: ["--bindIP", "192.0.0.2", "--remoteLoggerIP", "192.0.0.1",
46.                 "--remoteLoggerPort", "5000", "--masterIP", "192.0.0.1",
47.                 "--masterPort", "5001", "--containerName", "TempContainerName"]
48.          resources: {}
49.          volumeMounts:
50.          - mountPath: /var/run/docker.sock
51.            name: fogbus2-actor-hostpath0
52.        restartPolicy: Always
53.        serviceAccountName: ""
54.        nodeName: worker01
55.        hostNetwork: true
56.        volumes:
57.        - hostPath:
58.            path: /var/run/docker.sock
59.          name: fogbus2-actor-hostpath0
60. status: {}
```

In the communication design of the FogBus2 architecture, the initialization of components requires the binding of the host IP address, which will be used to pass information between components. For example, when a master component is created, the IP address of the host will be passed in as a required parameter. Although the IP address bound to master will also be passed as a necessary parameter to create the actor component, as the FogBus2 architecture has generic functions that will be used by multiple types or all components, the master component will still send its bound host IP address to the actor and tell it to return the information to this address. When the IP address bound by the master component is not the same as the IP address told by the master component, communication can be a problem. When the FogBus2 architecture is deployed using Docker Compose, communication between the components is smooth because the containers are running directly on the host. However, when the FogBus2 architecture is started in K3s, communication between the components does not work properly, due to the reason that containers are running in Pods and each Pod has its own IP address. Components cannot listen to the IP address of the host because by default, the Pod's network environment is separate from the host, which poses a challenge for the deployment of the FogBus2 architecture. To cope with this problem, we proposed the following three design models.

### 3.2.4.1 Host Network Mode

When starting FogBus2 components in a K3s cluster, instead of using the cluster's own network services, we use the host's network configuration directly. Specifically, we connect each Pod directly to the network of its host. In this case, the components in the FogBus2 framework can be bound directly to the host's network at initialization, and the IP address notified to the target component by each component is the same as the one configured by the target component at creation. Our experiments have successfully implemented this approach, and all components in the FogBus2 framework can communicate with each other successfully and work as a whole to provide services to the user. Figure 3.5 shows the schematic architecture of Design 1.

However, this design pattern sacrifices some of the functionality of K3s. When Pods are connected directly to the network environment where the hosts are located, the K3s controller will not be able to optimally manage all the containers within the cluster because these services are based on the K3s controller being able to have the highest level of access to the network services used by the Pods. If the Pods are on a VPN or WAN network, we will not be able to implement all the features of K3s. For example, the K3s controller cannot automatically increase the number of Pods because it cannot create a new IP address to assign to Pods in a VPN or WAN environment, and these functions need to be performed manually by the administrator.

### 3.2.4.2 Proxy Server

As the problem stems from a conflict between the communication design of the FogBus2 framework and the communication model between Pods in the K3s cluster, we can create a proxy server that defines the appropriate routing policies to receive and forward messages from different applications. When a FogBus2 component needs to send a message to another component, we import the message into the proxy server, which analyses the message to know the destination and sends it correctly to the IP address of the target component according to its internal routing policy. This approach bypasses the native communication model of the FogBus2 framework, and all communication between applications is done through the proxy server. Figure 3.6 shows the schematic architecture of Design 2.

There are two types of communication methods in the FogBus2 framework, proprietary methods and generic methods. The proprietary methods are used to communicate with fixed components, such as master and remote logger, whose IP addresses are configured and stored as global variables when most components are initialized. In contrast, the generic methods are used by all components and are called by components to transmit their IP addresses as part of the message for the target component to respond to. Therefore, in order to enable all components to send messages to the proxy server for processing, we need to change the source code of the FogBus2 framework so that all components are informed of the IP address of the proxy server at initialization, and to unify the two types of communication methods so that components will include information about the target in the message and send it to the proxy server. As a result, this design would involve a redesign of the communication model of the FogBus2 framework, which is not a good practice in the industry.

### 3.2.4.3 Environment Variable

In the K3s cluster, when the application is deployed, the cluster controller will automatically create a Pod to manage the container in which the application resides. However, in the YAML file, we can obtain the IP address of the created Pod when configuring the container information, which allows us to pass it in as an environment variable when initializing the components of the FogBus2 framework, so that the IP address bound to the application is the IP address of the Pod it is in and the component can transmit this address to the target component when communicating and receive a message back. Figure 3.7 shows the schematic architecture of Design 3.

However, in our experiments, we found that Pods on different nodes had problems communicating at runtime. We traced the flow of information transmitted and found that the reason for this was a conflict between the network services configured within the cluster to manage the Pods and the VPN connecting the cloud server to the edge virtual machines. The Pods were set up using their own unique network to assign IP addresses to the Pods and to communicate with each other, but the addresses cannot be recognized by the VPN set up on the nodes, which prevented the information from being transferred from the hosts. To solve this problem, we have proposed two solutions:

- **Solution 1:** K3s uses flannel as the Container Network Interface (CNI) by default. We can change the default network service configuration of the K3s cluster and override the default flannel interface with the Wireguard Ethereum Name Service.

- **Solution 2:** We can change the Wireguard settings to add the interface of the network service created by the K3s controller to the VPN profile to allow incoming or outgoing messages from a specific range of IP addresses.

Although we were unable to implement this design due to time constraints, we believe this design pattern is the best practice for deploying the FogBus2 framework into the K3s hybrid cluster, as the K3s controller is able to orchestrate the containers in the cluster using its own web services. Compared to the first design approach, this design implements all the functionality of K3s to orchestrate the containers in the cluster, and the controller can automatically increase and decrease the number of Pods to achieve reasonable scheduling of cluster resources; compared to the second design, this design does not require changes to the communication policy of the FogBus2 framework and saves the overhead of the proxy server in the cluster.

## 3.3 Experiment

We used Host Network Mode to deploy the FogBus2 framework to the K3s cluster to test the design and evaluate the response time. The application we adopted is to perform some simple mathematical calculations, as shown below:

```
1.  def Calculation(a, b, c):
2.      resultPart0 = a + b + c
3.      a += 1
4.      b += 1
5.      c += 1
6.      resultPart1 = a * a / (b * b + c * c)
7.      a += 1
```

```
8.      b += 1
9.      c += 1
10.     resultPart2 = 1 / a + 2 / b + 3 / c
11.     finalResult = resultPart0 + resultPart1 + resultPart2
12.     return resultPart0, resultPart1, resultPart2, finalResult
```

Moreover, Figure 3.4 shows that the FogBus2 framework has been successfully deployed in the K3s cluster.

```
NAME                                        READY   STATUS    RESTARTS   AGE    IP                NODE       NOMINATED NODE   READINESS GATES
fogbus2-remote-logger-c699c67c8-zbmsx       1/1     Running   0          143m   45.113.235.156    master     <none>           <none>
fogbus2-master-69858b5b8f-glrfc             1/1     Running   0          142m   45.113.235.156    master     <none>           <none>
fogbus2-actor1-5574767c65-fj6lt             1/1     Running   0          142m   45.113.232.199    worker01   <none>           <none>
fogbus-actor2-f7b7b9bfc-ctzxr               1/1     Running   0          142m   45.113.232.232    worker02   <none>           <none>
```

**Figure 3.4:** FogBus2 framework running in the K3s cluster

We have performed 10 experiments in total, and the system can return the correct results each time. We also conducted ten experiments on the native architecture of FogBus2 under the same network environment. The response time of all experiments is shown in Figure 3.5. The results show that when FogBus2 is run in K3s, the response time fluctuates between 30 milliseconds and 40 milliseconds, with an increase of 5 to 10 milliseconds compared to the native FogBus2 framework. In addition, the K3s cluster will also cause an increase in the jitter of the FogBus2 framework response time. However, considering the centralized resource management and scheduling and automated container health checks provided by K3s, we believe that these increases are acceptable.

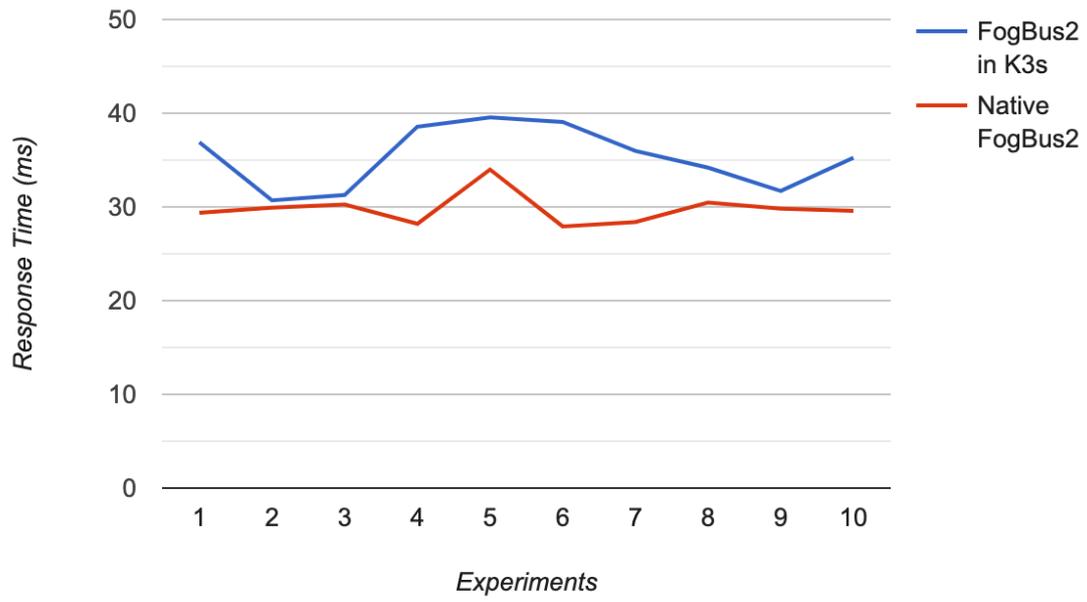

**Figure 3.5:** The impact of K3s cluster on FogBus2 in terms of response time

# Chapter 4  Conclusions and Future Work

This chapter concludes the project report and summarizes the proposed design. It also presents approaches to improve the current work in the future.

## 4.1   Conclusions

In this project report, we present feasible designs for implementing container orchestration techniques in cloud and edge/fog computing environments. The study solves the problem of connecting virtual instances due to IP addresses in different environments by building a hybrid environment with cloud nodes and edge devices using VPNs. Besides, the study proposes three design patterns for deploying the FogBus2 framework into the hybrid environment.

The Host Network Pattern connects the components of the cluster to the host network environment, using the native communication model of the FogBus2 framework by masking the internal network environment of the cluster, while avoiding the network conflict problems related to VPN. The Proxy Server Pattern redesigns the way components communicate with each other in the FogBus2 framework, creating a communication center to receive and forward messages from within the cluster, reducing the coupling between applications to a certain extent, and enhancing the scalability of the model. The Environment Variable Pattern retains the communication model of the FogBus2 framework and allows for the creation of independent network services within the cluster, enabling automatic control and scheduling of internal resources by the cluster controller. Compared to the original Fogbus2 framework, the new system enables resource limit control, health checks, and fault tolerance to cope with the ever-changing number and functionality of connected IoT devices. In addition, this work provides guidance and recommendations for the use of appropriate orchestration tools depending on the different computing environments and the potential challenges that exist.

## 4.2 Future Work

Future research can consider using different VPN tools or changing the internal network configuration of the cluster to achieve the best practice of integrating the FogBus2 framework in cloud and fog environments. In addition, future investigations can consider implementing different orchestration tools and software, including KubeEdge, Docker Swarm, and MicroK8s, to explore the impact of different integrated container orchestration technologies on the FogBus2 framework's processing of real-time and non-real-time IoT applications. The evaluated parameters include computing resources such as CPU and memory occupied, and the time it takes for the cluster to provide external services. Besides, a variety of scheduling policies can be implemented to automate application deployment and improve resource usage efficiency for clusters, ranging from heuristics to machine learning techniques [22, 23]. For example, scheduling Pods to nodes with smaller memory and CPU footprints to automatically balance the load on the cluster, or spreading replicative Pods across different nodes to avoid a crash on one node affecting the whole system.

# Bibliography

[1] J. Gubbi, R. Buyya, S. Marusic and M. Palaniswami, "Internet of Things (IoT): A vision, architectural elements, and future directions", Future Generation Computer Systems, vol. 29, no. 7, pp. 1645-1660, 2013. Available: 10.1016/j.future.2013.01.010.

[2] M. Aazam, I. Khan, A. Alsaffar and E. Huh, "Cloud of Things: Integrating Internet of Things and cloud computing and the issues involved", Proceedings of 2014 11th International Bhurban Conference on Applied Sciences & Technology (IBCAST) Islamabad, Pakistan, 14th - 18th January, 2014, 2014. Available: 10.1109/ibcast.2014.6778179.

[3] R. Buyya and S. Srirama, Fog and Edge Computing: Principles and Paradigms. Wiley, 2019, pp. 13-16.

[4] A. Dastjerdi and R. Buyya, "Fog Computing: Helping the Internet of Things Realize Its Potential", Computer, vol. 49, no. 8, pp. 112-116, 2016. Available: 10.1109/mc.2016.245.

[5] W. Shi, J. Cao, Q. Zhang, Y. Li and L. Xu, "Edge Computing: Vision and Challenges", IEEE Internet of Things Journal, vol. 3, no. 5, pp. 637-646, 2016. Available: 10.1109/jiot.2016.2579198.

[6] A. Celesti, D. Mulfari, M. Fazio, M. Villari and A. Puliafito, "Exploring Container Virtualization in IoT Clouds", 2016 IEEE International Conference on Smart Computing (SMARTCOMP), 2016. Available: 10.1109/smartcomp.2016.7501691.

[7] Z. Cai and R. Buyya, "Inverse Queuing Model based Feedback Control for Elastic Container Provisioning of Web Systems in Kubernetes", IEEE Transactions on Computers, pp. 1-1, 2021. Available: 10.1109/tc.2021.3049598.

[8] Q. Deng, M. Goudarzi and R. Buyya, "FogBus2: a lightweight and distributed container-based framework for integration of IoT-enabled systems with edge and cloud computing", Proceedings of the International Workshop on Big Data in Emergent Distributed Environments, 2021. Available: 10.1145/3460866.3461768.

[9] "K3s - Lightweight Kubernetes", Rancher Labs, 2021. [Online]. Available: https://rancher.com/docs/k3s/latest/en/.

[10] M. Goudarzi, Q. Deng and R. Buyya, "Resource Management in Edge and Fog Computing using FogBus2 Framework", 2021. Available: arXiv:2108.00591v1.

[11] "What is Kubernetes?", Kubernetes, 2021. [Online]. Available: https://kubernetes.io/docs/concepts/overview/what-is-kubernetes/

[12] "What is Kubernetes?", Red Hat, 2020. [Online]. Available: https://www.redhat.com/en/topics/containers/what-is-kubernetes.

[13] "Architecture", Rancher Labs, 2021. [Online]. Available: https://rancher.com/docs/k3s/latest/en/architecture/.

[14] "minikube start", minikube, 2021. [Online]. Available: https://minikube.sigs.k8s.io/docs/start/.

[15] "Using Minikube to Create a Cluster", Kubernetes, 2021. [Online]. Available: https://kubernetes.io/docs/tutorials/kubernetes-basics/create-cluster/cluster-intro/.

[16] M. Rodriguez and R. Buyya, "Container-based cluster orchestration systems: A taxonomy and future directions", Software: Practice and Experience, vol. 49, no. 5, pp. 698-719, 2018. Available: 10.1002/spe.2660.

[17] Z. Zhong and R. Buyya, "A Cost-Efficient Container Orchestration Strategy in Kubernetes-Based Cloud Computing Infrastructures with Heterogeneous Resources", ACM Transactions on Internet Technology, vol. 20, no. 2, pp. 1-24, 2020. Available: 10.1145/3378447.

[18] T. Goethals, F. De Turck and B. Volckaert, "FLEDGE: Kubernetes Compatible Container Orchestration on Low-Resource Edge Devices", Internet of Vehicles. Technologies and Services Toward Smart Cities, pp. 174-189, 2020. Available: 10.1007/978-3-030-38651-1_16.

[19] A. Pires, J. Simão and L. Veiga, "Distributed and Decentralized Orchestration of Containers on Edge Clouds", Journal of Grid Computing, vol. 19, no. 3, 2021. Available: 10.1007/s10723-021-09575-x.

[20] M. Alam, J. Rufino, J. Ferreira, S. Ahmed, N. Shah and Y. Chen, "Orchestration of Microservices for IoT Using Docker and Edge Computing", IEEE Communications Magazine, vol. 56, no. 9, pp. 118-123, 2018. Available: 10.1109/mcom.2018.1701233.

[21] D. Ermolenko, C. Kilicheva, A. Muthanna and A. Khakimov, "Internet of Things Services Orchestration Framework Based on Kubernetes and Edge Computing", 2021 IEEE Conference of Russian Young Researchers in Electrical and Electronic Engineering (ElConRus), 2021. Available: 10.1109/elconrus51938.2021.9396553.

[22] S. Agarwal, M. Rodriguez and R. Buyya, "A Reinforcement Learning Approach to Reduce Serverless Function Cold Start Frequency", 2021 IEEE/ACM 21st International Symposium on Cluster, Cloud and Internet Computing (CCGrid), 2021. Available: 10.1109/ccgrid51090.2021.00097

[23] M. Goudarzi, M. Palaniswami and R. Buyya, "A Distributed Deep Reinforcement Learning Technique for Application Placement in Edge and Fog Computing Environments", IEEE Transactions on Mobile Computing, pp. 1-1, 2021. Available: 10.1109/tmc.2021.3123165